\documentclass[12pt]{article}
\usepackage{amsmath,amsfonts,amssymb,amsthm,amsbsy,euscript}

\newtheorem{lemma}{Lemma}
\newtheorem*{criterion}{Criterion}
\theoremstyle{definition}
\newtheorem*{rem}{Remark}
\newtheorem*{conclusion}{Conclusion}

\title{On the 
variational noncommutative Poisson geometry}
\author{A.~V.~Kiselev}

\date{December 25, 2011}

\begin{document}
\maketitle


\begin{abstract}
We outline the notions and concepts of the calculus of variational
multivectors within the Poisson formalism over the spaces of infinite
jets of mappings from commutative (non)\/graded smooth manifolds to the
factors of noncommutative associative algebras over the equivalence under
cyclic permutations of the letters in the associative words. We state 
the basic properties of the variational Schouten bracket and derive an
interesting criterion for (non)\/commutative differential operators to be
Hamiltonian (and thus determine the (non)\/commutative Poisson 
structures).\ We place the noncommutative jet\/-\/bundle construction at hand in the
context of the quantum \mbox{string theory}.
\end{abstract}

\paragraph*{Introduction.}
In this brief communication we sum up, without giving any 
detailed proofs, the main notions and facts about the calculus of noncommutative 
variational multivectors in terms of the Schouten bracket, 
and we sketch the construction of the odd (homological) evolutionary vector fields, 
on the infinite jet spaces for maps of a smooth commutative ma\-ni\-fold~$M^n$ 
to the factor\/-\/algebra of a noncommutative associative
algebra~${\EuScript A}$ by the relations of the cyclic invariance
$a_1{\times} a_2\cdot a_3\sim a_1\cdot a_2{\times} a_3$ and
$a_1{\times} a_2\sim a_2{\times} a_1$ (see~\cite{KontsCyclic}),
where $\times$ is the ordered concatenation of the words $a_i$ 
written in the alphabet of~${\EuScript A}$ and $\cdot$~is the multiplication.

We formulate the basics of the theory over such noncommutative jet bundles;
this di\-rec\-ti\-on of research was pioneered in~\cite{OlverSokolov1998CMP}.
If, at the end of the day, the target algebra~${\EuScript A}$ is proclaimed
(graded-)\/commutative (and if it satisfies the ``smoothness'' assumptions),
we restore the stan\-dard, Gel'fand\/--\/Dorfman's calculus of variational 
multivectors~\cite{Olver,Topical}. Alternatively, under the shrinking of 
the source manifold~$M^n$, which may be our space\/-\/time, to a point 
(or by po\-stu\-la\-ting that the image of $M^n$
in~${\EuScript A}$ is a given element whenever the map $M^n\to{\EuScript A}$ is 
con\-stant), we reproduce the
noncommutative symplectic geometry of~\cite{KontsCyclic}.

This report is an immediate continuation of the 
review~\cite{Praha11} and we follow the notation and conventions thereof,
with the only replacement of~$\varSigma^n$ and~$\boldsymbol{q}$ there
by~$M^n$ and~$\boldsymbol{a}$ here, respectively.
A~proper substantiation and the discussion in more detail, 
with the transcript 
of this theory as the $\hbar$-\/linear slice 
of the scattering equation $({\mid}1\rangle\cdot{\mid}2\rangle)\cdot{\mid}3\rangle=
{\mid}1\rangle\cdot({\mid}2\rangle\cdot{\mid}3\rangle)$ in the
full quantum  geometry~\cite{KontsevichFormality},
will be the subject of the subsequent publication.



This paper is structured as follows. We first rephrase the notion of the variational 
co\-tan\-gent superbundle (see~\cite{Topical,Praha11} and references therein) 
in the noncommutative setup and formulate the definition of the noncommutative 
Schouten bracket~$[\![\,,\,]\!]$ as the odd Poisson bracket. We relate the odd 
evolutionary vector fields~$\smash{\boldsymbol{Q}^{\boldsymbol{\xi}}}$ to the operations
$[\![\boldsymbol{\xi},\cdot]\!]$, where $\boldsymbol{\xi}$~is a noncommutative 
variational multivector. In these terms, we debate the hamletian ``presence'' 
or ``absence'' of the Leibniz rule for the Schouten bracket~$[\![\,,\,]\!]$. 
We affirm 
the shifted\/-\/graded skew\/-\/symmetry of~$[\![\,,\,]\!]$ and 
directly verify the Jacobi identity, which stems from the usual
Leibniz rule for the derivations $\smash{\boldsymbol{Q}^{\boldsymbol{\xi}}}$
acting on the bracket $[\![\boldsymbol{\eta},\boldsymbol{\omega}]\!]$. 
Later on, 
we focus on the noncommutative variational Poisson bivectors $\boldsymbol{\pi}$ 
such that $[\![\boldsymbol{\pi},\boldsymbol{\pi}]\!]=0$ or, 
equivalently, $\tfrac{1}{2}\smash{\bigl(\boldsymbol{Q}^{\boldsymbol{\pi}}\bigr)^2}=0$.
We derive an interesting criterion for a (non)\/commutative linear operator~$A$ to be
Hamiltonian (resp., for the bivector $\boldsymbol{\pi}=
\tfrac{1}{2}\langle\boldsymbol{b},A(\boldsymbol{b})\rangle$
to satisfy $[\![\boldsymbol{\pi},\boldsymbol{\pi}]\!]=0$).

\paragraph*{1. Noncommutative jets.}
Let $M^n$ be a smooth oriented $\mathbb{R}$-\/manifold and $\boldsymbol{x}\in M^n$ 
be a point. 
Let ${\EuScript A}$ be a noncommutative associative 
algebra of dimension~$m$; 
denote by $\boldsymbol{a}$ a basis in~${\EuScript A}$.
Consider the maps $M^n\to{\EuScript A}$ and construct the infinite jet space
$J^\infty(M^n\to{\EuScript A})\mathrel{{=}{:}} J^\infty(\pi^{\text{nC}})$,
see~\cite{Praha11}.
Denote by $(a\cdot)$ and $(\cdot a)$ the operators of left-{} and right\/-\/multiplication
by a word $a\in{\EuScript A}$ that is always read from left to right. The total
derivative w.r.t.\ $x^i$, $1\leqslant i\leqslant n$, on $J^\infty(\pi^{\text{nC}})$
is $\vec{\mathrm{d}}/\mathrm{d}x^i=\vec{\partial}/\partial x^i +\sum_{|\sigma|\geqslant0}
\boldsymbol{a}_{\sigma+1_i} \vec{\partial}/\partial\boldsymbol{a}_\sigma$;
the evolutionary derivation $\smash{\partial^{(\boldsymbol{a})}_\varphi}=
\smash{\sum_{|\sigma|\geqslant0}
\bigl(\smash{\vec{\mathrm{d}}^{|\sigma|}}/\mathrm{d}\boldsymbol{x}^\sigma\bigr)
(\varphi)\cdot\smash{\vec{\partial}}/\partial\boldsymbol{a}_\sigma}$ 
acts from the left by the 
Leibniz rule.
Denote by $\bar{\Lambda}^n(\pi^{\text{nC}})$ the 
$C^\infty(J^\infty(\pi^{\text{nC}}))$-\/module of horizontal forms of the
highest ($n$-th) degree and by $\bar{H}^n(\pi^{\text{nC}})$ the respective 
cohomology   
w.r.t.\ the horizontal differential $\bar{\mathrm{d}}
=\sum_{i=1}^n \mathrm{d}x^i\cdot \vec{\mathrm{d}}/\mathrm{d}x^i$;
the Cartan differential on $J^\infty(\pi^{\text{nC}})$ is $\mathrm{d}_{\mathcal{C}}=
 \mathrm{d}_{\text{dR}}-\bar{\mathrm{d}}$.
Denote by $\langle\,,\,\rangle$ the $\bar{\Lambda}^n(\pi^{\text{nC}})$-\/valued coupling
between the spaces of variational covectors $\boldsymbol{p}$ and evolutionary vectors
$\smash{\partial^{(\boldsymbol{a})}_\varphi}$. By default, we pass to the cohomology and,
using the integration by parts,
normalize the (non)\/commutative covectors as follows, $\boldsymbol{p}\bigl(
\boldsymbol{x},[\boldsymbol{a}]\bigr)=
\smash{\sum_{j=1}^m} \langle\text{word}\rangle\cdot\mathrm{d}_{\mathcal{C}}a^j\cdot
\langle\text{word}\rangle\cdot\mathrm{d}\boldsymbol{x}$; one then can freely push
$\mathrm{d}_{\mathcal{C}}\boldsymbol{a}$ left-{} or rightmost using the cyclic
invariance. (Due to the structure of $\smash{\partial^{(\boldsymbol{a})}_\varphi}$,
the value $\langle\boldsymbol{p},\varphi\rangle$ is well\/-\/defined irrespective
of the normalization in~$\boldsymbol{p}$.)

Let $A\colon\boldsymbol{p}\to\varphi$ be a Noether noncommutative linear matrix operator
in total derivatives. 
The 
adjoint operator $\vec{A}^\dagger$ is defined from the equality
$\langle\boldsymbol{p}_1,A(\boldsymbol{p}_2)\rangle=
\langle\boldsymbol{p}_2,\vec{A}^\dagger(\boldsymbol{p}_1)\rangle$, in which we first
integrate by parts and then transport 
the even covector $\boldsymbol{p}_2$ around the circle.

\begin{lemma}\label{LiftP}
For each $\dot{\boldsymbol{a}}=\varphi$\textup{,} the induced velocity $\dot{\boldsymbol{p}}=
\mathrm{L}_{\partial^{(\boldsymbol{a})}_\varphi}(\boldsymbol{p})$ equals
$\dot{\boldsymbol{p}}=\partial^{(\boldsymbol{a})}_\varphi(\boldsymbol{p})+
(\boldsymbol{p})\smash{\overleftarrow{ \bigl(\ell^{(\boldsymbol{a})\,\dagger}_\varphi
\bigr)}}$\textup{,}
where ${\ell^{(\boldsymbol{a})\,\dagger}_\varphi}$ is the adjoint to the 
linearization\textup{,}
which is $\ell^{(\boldsymbol{a})}_\varphi(\boldsymbol{\alpha})=
\tfrac{\mathrm{d}}{\mathrm{d}\varepsilon}{\bigr|}_{\varepsilon=0} 
\varphi\bigl(\boldsymbol{x},[\boldsymbol{a}+\varepsilon\boldsymbol{\alpha}]\bigr)$.
\end{lemma}

\paragraph*{2. Noncommutative multivectors.}
The covectors $\boldsymbol{p}\bigl(\boldsymbol{x},[\boldsymbol{a}]\bigr)$ were even.
We reverse their parity, $\Pi\colon\boldsymbol{p}\mapsto
\boldsymbol{b}\bigl(\boldsymbol{x},[\boldsymbol{a}]\bigr)$, 
preserving the topology but endowing the space of differential functions $f$
that depend on~$\boldsymbol{b}$ with a new ring structure: now, each $f$ is 
polynomial in fi\-ni\-te\-ly many derivatives of $\boldsymbol{b}$. 
Next, we consider the noncommutative variational cotangent superspace
$\overline{J^\infty}\bigl(\Pi\widehat{\pi}^{\text{nC}}_\pi\bigr)=
J^\infty(\Pi\widehat{\pi}^{\text{nC}})\mathbin{{\times}_{M^n}} J^\infty(\pi^{\text{nC}})$,
see~\cite{Olver,Topical} and~\cite{Praha11}. In effect, we declare that 
$\boldsymbol{b}$,\ $\boldsymbol{b}_{\boldsymbol{x}}$,\ %
$\boldsymbol{b}_{\boldsymbol{x}\boldsymbol{x}}$,\ $\ldots$,\ $\boldsymbol{b}_{\tau}$ are
the extra, odd jet variables on top of the old, even~$\boldsymbol{a}_{\sigma}$'s.
The total derivatives $\vec{\mathrm{d}}/\mathrm{d}x^i$ obviously lift onto
$\overline{J^\infty}\bigl(\Pi\widehat{\pi}^{\text{nC}}_\pi\bigr)$ as well as
$\bar{\mathrm{d}}$ that yields the cohomology 
$\bar{H}^n\bigl(\Pi\widehat{\pi}^{\text{nC}}_\pi\bigr)=
 \bar{\Lambda}^n\bigl(\Pi\widehat{\pi}^{\text{nC}}_\pi\bigr)\bigr/
  (\text{im}\,\bar{\mathrm{d}})$.
The two components of the evolutionary vector fields $\boldsymbol{Q}=
\smash{\partial^{(\boldsymbol{a})}_{\varphi^{\boldsymbol{a}}}+
\partial^{(\boldsymbol{b})}_{\varphi^{\boldsymbol{b}}}}$ now begin with
$\dot{\boldsymbol{a}}=
\varphi^{\boldsymbol{a}}\bigl(\boldsymbol{x},[\boldsymbol{a}],[\boldsymbol{b}]\bigr)$ and
$\dot{\boldsymbol{b}}=
\varphi^{\boldsymbol{b}}\bigl(\boldsymbol{x},[\boldsymbol{a}],[\boldsymbol{b}]\bigr)$,
c.f.~Lemma~\ref{LiftP}.

The definition of noncommutative variational $k$-\/vectors, their evaluation on
$k$ covectors, the definition of the noncommutative variational antibracket, and its
inductive calculation are 
two pairs of distinct concepts.
A noncommutative $k$-\/vector~$\boldsymbol{\xi}$, $k\geqslant0$, 
is a cohomology class in $\bar{H}^n\bigl(\Pi\widehat{\pi}^{\text{nC}}_\pi\bigr)$
whose density is $k$-\/linear in the odd $\boldsymbol{b}$'s or their derivatives.
Each $\boldsymbol{\xi}$ 
can be normalized to $\boldsymbol{\xi}=\langle\boldsymbol{b},
A(\boldsymbol{b},\ldots,\boldsymbol{b})\rangle/k!$, where the noncommutative total 
differential operator $A$ depends on $(k-1)$ odd entry and may 
have $\boldsymbol{a}$-\/dependent coefficients. Integrating by parts and pushing the 
letters of the word~$\boldsymbol{\xi}$ along the circle, we infer that 
$\langle\boldsymbol{b}_1,A(\boldsymbol{b}_2,\ldots,\boldsymbol{b}_k)\rangle
\smash{{\bigr|}_{\boldsymbol{b}_i\mathrel{{:}{=}}\boldsymbol{b}}}=
\langle\boldsymbol{b}_2,A^\dagger_\circlearrowright(\boldsymbol{b}_3,\ldots,
\boldsymbol{b}_k,\boldsymbol{b}_1)\rangle
\smash{{\bigr|}_{\boldsymbol{b}_i\mathrel{{:}{=}}\boldsymbol{b}}}$;
note that, each time an odd variable $\boldsymbol{b}_\tau$ reaches a marked point
$\boldsymbol{\infty}$ on the circle, it counts the $k-1$ 
other odd variables whom it overtakes and reports the sign $(-)^{k-1}$
(in particular, $A^\dagger_\circlearrowright=A^\dagger=-A$ if~$k=2$).
The value of the $k$-\/vector $\boldsymbol{\xi}$ on $k$ arbitrary 
covectors $\boldsymbol{p}_i$ is $\boldsymbol{\xi}(\boldsymbol{p}_1,\ldots,
\boldsymbol{p}_k)=\sum_{s\in S_k}(-)^{|s|}\langle\boldsymbol{p}_{s(1)},
A(\boldsymbol{p}_{s(2)},\ldots,\boldsymbol{p}_{s(k)})\rangle/k!$; 
we emphasize that we
shuffle the arguments but never swap their slots, which are built into the cyclic
word~$\boldsymbol{\xi}$.  

\paragraph*{3. Noncommutative Schouten bracket.}
The concatenation~$\times$ of densities of two multivectors provides an ill\/-\/defined
product 
in $\bar{H}^n\bigl(\Pi\widehat{\pi}^{\text{nC}}_\pi\bigr)$,
where the genuine multiplication
is the noncommutative antibracket.
We 
fix the Dirac ordering 
$\delta\boldsymbol{a}\wedge\delta\boldsymbol{b}$ over 
each $\boldsymbol{x}$ in $\overline{J^\infty}\bigl(\Pi\widehat{\pi}^{\text{nC}}_\pi\bigr)
\to M^n$; 
note that $\delta\boldsymbol{a}$ is a covector and $\delta\boldsymbol{b}$ is
an odd vector so that their coupling equals $+1\cdot\mathrm{d}\boldsymbol{x}$.
The noncommutative variational Schouten bracket of two multivectors $\boldsymbol{\xi}$
and $\boldsymbol{\eta}$ is $[\![\boldsymbol{\xi},\boldsymbol{\eta}]\!]=
\langle\overrightarrow{\delta}\!\boldsymbol{\xi}\wedge
\overleftarrow{\delta}\!\boldsymbol{\eta}\rangle$.\ In coordinates,\ this yields
$[\![\boldsymbol{\xi},\boldsymbol{\eta}]\!]=
\bigl[\overrightarrow{\delta}\!\boldsymbol{\xi}/\delta\boldsymbol{a}\,\cdot
\overleftarrow{\delta}\!\boldsymbol{\eta}/\delta\boldsymbol{b}\,-
\overrightarrow{\delta}\!\boldsymbol{\xi}/\delta\boldsymbol{b}\,\cdot
\overleftarrow{\delta}\!\boldsymbol{\eta}/\delta\boldsymbol{a}\bigr]$, where (1)
all the derivatives are thrown off the variations $\delta\boldsymbol{a}$ and
$\delta\boldsymbol{b}$ via the integration by parts, then (2) the letters
$\boldsymbol{a}_\sigma$, $\boldsymbol{b}_\tau$, $\delta\boldsymbol{a}$, and
$\delta\boldsymbol{b}$, which are thread on the two circles 
$\delta\boldsymbol{\xi}$ and $\delta\boldsymbol{\eta}$,
spin 
along these rosaries 
so that the variations $\delta\boldsymbol{a}$ and $\delta\boldsymbol{b}$ match 
in all possible combinations, and finally, (3) the variations 
$\delta\boldsymbol{a}$ and $\delta\boldsymbol{b}$ detach from the circles and couple,
while the loose ends of the two remaining open strings 
join 
and form the new circle. 

The Schouten bracket is shifted\/-\/graded skew\/-\/symmetric:
if $\boldsymbol{\xi}$ is a $k$-\/vector and $\boldsymbol{\eta}$ 
an $\ell$-\/vector, then $[\![\boldsymbol{\xi},\boldsymbol{\eta}]\!]=
-(-)^{(k-1)(\ell-1)}[\![\boldsymbol{\eta},\boldsymbol{\xi}]\!]$.

\paragraph*{4. Is $[\![\,,\,]\!]$ a bi\/-\/derivation?}
In the notation of \S3, define the evolutionary vector field 
$\smash{\boldsymbol{Q}^{\boldsymbol{\xi}}}$ on
$\overline{J^\infty}\bigl(\Pi\widehat{\pi}^{\text{nC}}_\pi\bigr)$
by the rule $\smash{\boldsymbol{Q}^{\boldsymbol{\xi}}}(\boldsymbol{\eta})=
[\![\boldsymbol{\xi},\boldsymbol{\eta}]\!]$, whence
$\smash{\boldsymbol{Q}^{\boldsymbol{\xi}}}=
-\partial^{(\boldsymbol{a})}_{\vec{\delta}\boldsymbol{\xi}/\delta\boldsymbol{b}}
+\partial^{(\boldsymbol{b})}_{\vec{\delta}\boldsymbol{\xi}/\delta\boldsymbol{a}}$.
The normalization $\boldsymbol{\xi}=\bigl\langle\boldsymbol{b},
A(\boldsymbol{b},\ldots,\boldsymbol{b}\bigr\rangle/k!$ determines
$\smash{\boldsymbol{Q}^{\boldsymbol{\xi}}}=-(-)^{k-1}\tfrac{1}{(k-1)!}
\partial^{(\boldsymbol{a})}_{A(\boldsymbol{b},\ldots,\boldsymbol{b})}
+(-)^{k-1}\tfrac{1}{k!}\smash{\partial^{(\boldsymbol{b})}_{
\vec{\ell}^{\,(\boldsymbol{a})\,\dagger}_{A(\boldsymbol{b}_2,\ldots,\boldsymbol{b}_k)}
 (\boldsymbol{b}_1)}} \smash{{\bigr|}_{\boldsymbol{b}_i\mathrel{{:}{=}}\boldsymbol{b}}}$;
e.g., $\boldsymbol{Q}^{\frac{1}{2}\langle\boldsymbol{b},A(\boldsymbol{b})\rangle}=
\partial^{(\boldsymbol{a})}_{A(\boldsymbol{b})}-\tfrac{1}{2}
\partial^{(\boldsymbol{b})}_{\vec{\ell}^{\,(\boldsymbol{a})\,\dagger}_{A(\boldsymbol{b})}
 (\boldsymbol{b})}$, see~\cite{Topical,Praha11}.

Freeze the coordinates, fix the volume form on $M^n$, and choose any representatives
$\boldsymbol{\xi}$ and $\boldsymbol{\eta}$ of the cohomology classes in
$\bar{H}^n\bigl(\Pi\widehat{\pi}^{\text{nC}}_\pi\bigr)$. The derivation
$\smash{\boldsymbol{Q}^{\boldsymbol{\xi}}}$ acts on the word $\boldsymbol{\eta}$
by the graded Leibniz rule, inserting 
$\smash{\tfrac{\vec{\mathrm{d}}^{|\sigma|}}{\mathrm{d}\boldsymbol{x}^\sigma}}\bigl(
\boldsymbol{Q}^{\boldsymbol{\xi}}(\boldsymbol{q})\bigr)$ instead of each letter
$\boldsymbol{q}_\sigma$ (here $\boldsymbol{q}$ is $\boldsymbol{a}$ or $\boldsymbol{b}$).
Next, promote the letter $\boldsymbol{q}$ to the ze\-ro-{} or one\/-\/vector
$\boldsymbol{q}\cdot\mathrm{d}\boldsymbol{x}\in
\bar{\Lambda}^n\bigl(\Pi\widehat{\pi}^{\text{nC}}_\pi\bigr)$ and use the Leibniz rule
again to expand the entries $[\![\boldsymbol{\xi},\boldsymbol{q}]\!]=
(\boldsymbol{\xi})
\overleftarrow{\boldsymbol{Q}^{\boldsymbol{q}}}
$.
The bracket $[\![\,,\,]\!]\colon\bar{\Lambda}^n\bigl(\Pi\widehat{\pi}^{\text{nC}}_\pi\bigr)
\times\bar{\Lambda}^n\bigl(\Pi\widehat{\pi}^{\text{nC}}_\pi\bigr)\to
\bar{\Lambda}^n\bigl(\Pi\widehat{\pi}^{\text{nC}}_\pi\bigr)$ becomes 
a derivation in each argument.
However, the calculation of $[\![\boldsymbol{\xi},\boldsymbol{\eta}]\!]$ is ill\/-\/defined
if one permits the addition of $\bar{\mathrm{d}}$-\/exact terms (e.g., stemming from
the integration by parts) to the entries $[\![\boldsymbol{\xi},\boldsymbol{q}]\!]$.
Besides, the normalization of the final result is a must 
in order to let us compare any given multivectors; in fact, 
the usual commutator
of one\/-\/vectors is always transformed to $\bigl\langle\boldsymbol{b},
-\bigl(\partial^{(\boldsymbol{a})}_{\varphi_1}(\varphi_2)-
 \partial^{(\boldsymbol{a})}_{\varphi_2}(\varphi_1)\bigr)\bigr\rangle=
[\![\langle\boldsymbol{b},\varphi_1\rangle,\langle\boldsymbol{b},\varphi_2\rangle]\!]$,
with no derivatives falling on~$\boldsymbol{b}$. At this point, the Leibniz rule
for $[\![\,,\,]\!]$ is in general irreparably lost.

The conceptual equality $\bigl[\boldsymbol{Q}^{\boldsymbol{\xi}},
\boldsymbol{Q}^{\boldsymbol{\eta}}\bigr]=\boldsymbol{Q}^{[\![\boldsymbol{\xi},
\boldsymbol{\eta}]\!]}$, with the graded commutator in its l.-h.s.\ and the noncommutative
variational Schouten bracket in the r.-h.s., proves that the Leibniz rule
$\boldsymbol{Q}^{\boldsymbol{\xi}}\bigl([\![\boldsymbol{\eta},\boldsymbol{\omega}]\!]\bigr)
=[\![\boldsymbol{Q}^{\boldsymbol{\xi}}(\boldsymbol{\eta}),\boldsymbol{\omega}]\!]
+(-)^{(k-1)(\ell-1)}[\![\boldsymbol{\eta},
\boldsymbol{Q}^{\boldsymbol{\xi}}(\boldsymbol{\omega})]\!]$, where
$\boldsymbol{\omega}\in\bar{H}^n\bigl(\Pi\widehat{\pi}^{\text{nC}}_\pi\bigr)$,
is the Jacobi identity $[\![\boldsymbol{\xi},[\![\boldsymbol{\eta},\boldsymbol{\omega}]\!]
 ]\!]=[\![ [\![\boldsymbol{\xi},\boldsymbol{\eta}]\!],\boldsymbol{\omega}]\!]
+(-)^{(k-1)(\ell-1)}[\![\boldsymbol{\eta},[\![\boldsymbol{\xi},\boldsymbol{\omega}]\!] 
]\!]$ for the bracket~$[\![\,,\,]\!]$.

\paragraph*{5. Noncommutative Poisson brackets.}
Each skew\/-\/adjoint noncommutative linear total differential operator
$A\colon\boldsymbol{p}\mapsto\dot{\boldsymbol{a}}=\varphi$ yields the bivector
$\boldsymbol{\pi}=\tfrac{1}{2}\langle\boldsymbol{b},A(\boldsymbol{b})\rangle$.
Let $\mathcal{H}_1$,\ $\mathcal{H}_2$,\ $\mathcal{H}_3$ be zero\/-\/vectors, i.e., 
$\mathcal{H}_i=\bigl[h_i(\boldsymbol{x},[\boldsymbol{a}])\,
\mathrm{d}\boldsymbol{x}\bigr]$. By definition, put $\{\mathcal{H}_i,\mathcal{H}_j\}_A
\mathrel{{:}{=}}\boldsymbol{\pi}\bigl(\smash{\vec{\delta}}
\mathcal{H}_i/\delta\boldsymbol{a},
\smash{\vec{\delta}}\mathcal{H}_j/\delta\boldsymbol{a}\bigr)$, which equals
$\bigl\langle\vec{\delta}\mathcal{H}_i/\delta\boldsymbol{a},
 A\bigl(\vec{\delta}\mathcal{H}_j/\delta\boldsymbol{a}\bigr)\bigr\rangle=
\partial^{(\boldsymbol{a})}_{A(\vec{\delta}\mathcal{H}_j/\delta\boldsymbol{a})}
 (\mathcal{H}_i)\pmod{\text{im}\,\bar{\mathrm{d}}}$. The bra\-cket $\{\,,\,\}_A$ is
bilinear and skew\/-\/symmetric (but it does not restrict as a bi\/-\/derivation
to the co\-ho\-mo\-lo\-gy w.r.t.\ $\bar{\mathrm{d}}$); it becomes Poisson if it
satisfies the Jacobi identity $\sum_{\circlearrowright}\bigl\{ \{\mathcal{H}_1,
 \mathcal{H}_2\}_A,\mathcal{H}_3\bigr\}_A=0$, which also is $\sum_{s\in S_3} (-)^{|s|} 
\smash{\partial^{(\boldsymbol{a})}_{A(\vec{\delta}\mathcal{H}_{s(3)}/\delta\boldsymbol{a})}}
\bigl(\tfrac{1}{2}\bigl\langle\smash{\vec{\delta}}\mathcal{H}_{s(1)}/\delta\boldsymbol{a},
A\bigl(\smash{\vec{\delta}}\mathcal{H}_{s(2)}/\delta\boldsymbol{a}\bigr)\bigr\rangle\bigr)=0$.
The tempting notation $\partial^{(\boldsymbol{a})}_{A(\boldsymbol{b})}(\boldsymbol{\pi})
\bigl(\smash{\bigotimes^3 \vec{\delta}}\mathcal{H}_i/\delta\boldsymbol{a}\bigr)=0$ 
is illegal by Lemma~\ref{LiftP} that forbids us  
to set $\dot{\boldsymbol{p}}\equiv0$ at will 
so that $\partial^{(\boldsymbol{a})}_{A(\boldsymbol{b})}$ would be ill\/-\/defined
on $\overline{J^\infty}\bigl(\Pi\widehat{\pi}^{\text{nC}}_\pi\bigr)$. 
Instead, we step farther and reach the clas\-si\-cal noncommutative master equation
$\smash{\boldsymbol{Q}^{\boldsymbol{\pi}}}(\boldsymbol{\pi})=
[\![\boldsymbol{\pi},\boldsymbol{\pi}]\!]=0$ upon the Poisson structures.
\begin{criterion}[\textmd{see~\cite{Praha11,Galli10} on $\smash{\tfrac{1}{2}
\bigl(\boldsymbol{Q}^{\boldsymbol{\pi}}\bigr)^2=0}$}]
A skew\/-\/adjoint \textup{(}non\textup{)}\/commutative 
linear matrix operator $A\colon\boldsymbol{p}\mapsto
\dot{\boldsymbol{a}}=\varphi$ in total derivatives is Hamiltonian
--i.e.\textup{,} the bivector $\boldsymbol{\pi}$ is Poisson-- if and only if 
its image is involutive\textup{:} 
$[\text{\textup{im}}\,A,\text{\textup{im}}\,A]\subseteq\text{\textup{im}}\,A$.
\end{criterion}

\begin{rem}
The construction of $[\![\,,\,]\!]$ in \S3 
is the standard string theory's pair of pants
$\varSigma_{{\mid}i\rangle}\cdot\varSigma_{{\mid}j\rangle}\mapsto
\varSigma_{{\mid}i\cdot j\rangle}$ flying \emph{over} the Minkowski 
space\/-\/time~$M^{3,1}$; the dimension reduction is not required.\ 
Neither the diameters of the circles
$\varSigma\simeq\mathbb{S}^1$ that carry the words
nor their stret\-ching 
or oscillations on them, but it is the information about the cyclic order 
that matters.
\end{rem}

\begin{conclusion}
The linking of the 
words into circles is natural for string theory, so the 
calculus of noncommutative variational multivectors may give new insights in it.
Yet, for the alphabet of ${\EuScript A}$ to fully 
depict 
the quantum world, we must quantize both the algebra and the Poisson brackets $\{\,,\,\}_A$
to their $\hbar$-\/deformations ${\EuScript A}[\hbar]$ and 
$\{\,,\,\}^{[\hbar]}_{A[\hbar]}$.\ This 
will be discussed elsewhere.
\end{conclusion}

\paragraph*{Acknowledgements.}
The author is grateful to the Organizing committee
of the international workshop SQS'11 
`Supersymmetry and Quantum Symmetries'
(July~18--23, 2011; JINR Dubna, Russia)
for a welcome and warm atmosphere during the meeting.
The author thanks B.~A.~Dubrovin, E.~A.~Ivanov,
S.~O.~Krivonos, 
J.~W.~van de~Leur, 
P.~J.~Olver, and V.~V.~Sokolov
for helpful discussions and stimulating remarks. 

This research was supported in part by NWO~VENI 
grant~639.031.623 (Utrecht) and JBI~RUG 
project~103511 (Groningen). 
A~part of this research was done while the author was visiting at 
the $\smash{\text{IH\'ES}}$ (Bures\/-\/sur\/-\/Yvette); 
the financial support and hospitality of this institution are gratefully acknowledged.


\begin{thebibliography}{7}

\bibitem{KontsCyclic}
Kontsevich M.
Formal (non)\/commutative symplectic geometry //
The Gel'fand Mathematical Seminars, 1990-1992.
Boston, MA: Birk\-h\"au\-ser, 1993. P.~173--187.

\bibitem{OlverSokolov1998CMP}
{Olver P. J., Sokolov V. V.} 
Integrable evolution equations on associative algebras //
{Commun.\ Math.\ Phys.} 1998. V.~{193}, n.~2. P.~245--268.
%

\bibitem{Olver}
{Olver P. J.} 
{Applications of Lie groups to differential equations}.
Grad.\ Texts in Math.\ V.~{107} (2nd ed.). 
N.Y.: Springer\/--\/Verlag, 1993. 

\bibitem{Topical}
{Krasil'shchik I., Verbovetsky A.} 
Geometry of jet spaces and integrable systems //
{J.~Geom.\ Phys.} 2011. V.~{61}. P.~1633--1674.

\bibitem{Praha11}
{Kiselev A. V.}
Homological evolutionary vector fields in
Korteweg\/--\/de Vries, Liouville, Maxwell,
and several other models.
Preprint 
$\smash{\text{IH\'ES}}$/M/11/26. 2011. 20~p.

\bibitem{KontsevichFormality}
Kontsevich M.
Deformation quantization of Poisson manifolds.~I //
{Lett.\ Math.\ Phys.} 2003. V.~66, n.~3.
P.~157--216.\ \texttt{arXiv:q-alg/9709040}

%


\bibitem{Galli10}
{Kiselev A. V., van de Leur J. W.} 
Variational Lie algebroids and homological evolutionary vector fields //
{Theor.\ Math.\ Phys.} 2011. V.~167, n.3. P.~772--784.

\end{thebibliography}
\end{document}